\documentclass[superscriptaddress,showpacs,prb,twocolumn]{revtex4}

\usepackage{amsmath}
\usepackage{amssymb}
\usepackage{epsfig}
\usepackage{graphicx}
\usepackage{wasysym}
\usepackage{color}
\newcommand \be{\begin{equation}}
\newcommand \ee{\end{equation}}
\newcommand \bes{\begin{equation*}} 
\newcommand \ees{\end{equation*}}
\newcommand \bea{\begin{eqnarray}}
\newcommand \eea{\end{eqnarray}}
\newcommand \beas{\begin{eqnarray*}} 
\newcommand \eeas{\end{eqnarray*}}
\newcommand \bfg{\begin{figure}}
\newcommand \efg{\end{figure}}
\newcommand \bfgs{\begin{figure*}} 
\newcommand \efgs{\end{figure*}}
\newcommand \bwt{\begin{widetext}}
\newcommand \ewt{\end{widetext}}
\renewcommand{\v}[1]{{\bf #1}}

\newcommand{\Fig}[1]{Fig.~\ref{#1}}

\newcommand \nd{{\vphantom{\dagger}}} 

\newcommand \tbl[1]{Table~\ref{#1}}

\newcommand \vecr{{\bf r}}
\newcommand \vk{{\bf k}}

\newcommand \etal{{\it et al.}}
\newcommand \im{i}

\newcommand \spm{$s_\pm$}

\begin{document}
\title{Nodes in the Gap Function of LaFePO, 
the Gap Function of the Fe(Se,Te) Systems, 
and the STM Signature of the $s_{\pm}$ Pairing}
\author{Fa Wang}
\affiliation{Department of Physics, Massachusetts Institute of Technology, Cambridge, Massachusetts 02139, USA}
\author{Hui Zhai}
\affiliation{Institute for Advanced Study, Tsinghua University, Beijing 100084, China}
\author{Dung-Hai Lee}
\affiliation{Department of Physics, University of California at Berkeley, Berkeley, California 94720, USA}
\affiliation{Materials Sciences Division, Lawrence Berkeley National Laboratory, Berkeley, California 94720, USA}

\date{\today}

\begin{abstract}
We reiterate, in more details, our previous proposal of using quasi-particle 
interference to determine the pairing form factor in iron-based superconductors. 
We also present our functional renormalization group(FRG) results on LaFePO 
and Fe(Se,Te) superconductors. In particular we found that 
the leading pairing channel in LaFePO is nodal $s_{\pm}$, 
with nodes on electron Fermi surfaces. 
For Fe(Se,Te) system we found fully gapped $s_{\pm}$ pairing, 
with substantial gap anisotropy on electron Fermi surfaces, 
and large gap is concentrated in regions with dominant $xy$ orbital character. 
We further fit the form factor obtained by FRG to real space orbital 
basis pairing picture, which shows more clearly the differences between 
different iron-based superconductors. 
\end{abstract}
\maketitle

In the midst of many remaining issues of the iron-based superconductors, 
the ``pairing symmetry'' has attracted a lot of attention. 
From the theoretical side, the answer converged rather quickly to the {\spm} 
form factor\cite{Mazin, Kuroki, LiJX, HuJP, FW, Chubukov, ZhangFC}. 
However, on the experimental side conflicting evidences exist for 
the presence of nodes\cite{nodes} and full gap\cite{fullgap}.  
Note that we use ``form factor'' rather than ``pairing-symmetry'' to describe {\spm}. 
This is because from the symmetry point of view there is no difference between 
the {\spm} and the usual $s$-wave pairing. 
Indeed, under the action of the crystal point group, both transform as 
the identity (trivial) representation. This is in marked contrast with 
the $d_{x^2-y^2}$ pairing symmetry of the cuprates, which transforms as 
a distinct irreducible representation upon the point group operations. 
Thus while there is sharp (symmetry) distinction between the $d_{x^2-y^2}$ 
and the usual $s$-wave pairing, such distinction does not exist 
between {\spm} and $s$. This is why phase sensitive measurements probing 
the relative phase of the superconducting order parameters residing in regions with 
different spatial orientation are ideal to rule in or rule out 
the $d_{x^2-y^2}$ pairing\cite{Tsuei,Harlingen}, 
while they can not definitively prove or disprove the {\spm} form factor\cite{Moler}.

Nonetheless at this point there exist two types of proposed experiment 
which can address whether the gap function on the electron and hole pockets 
are indeed of opposite sign. Both of them provide information about 
the relative (spinor) phase of the quasiparticle wave functions on 
the electron and hole pockets.  One of them is the detection of 
neutron resonance mode in the superconducting state. As suggested in 
Ref.~\cite{Maier,Korshunov}, if the pairing form factor is {\spm} 
one expects neutron resonance peaks at momenta $(\pm\pi,0),\,(0,\pm\pi)$ 
to occur in the superconducting state. 
The energy location of the peaks is less than the sum of the minimum gaps on 
the electron and hole pockets. 
Interestingly the resonance peak has been observed 
\cite{Lumsden,ChiS,LiS,Inosov,Zhao}.

\begin{figure*}[tbp]
\begin{center}
\includegraphics[angle=0,scale=0.5]{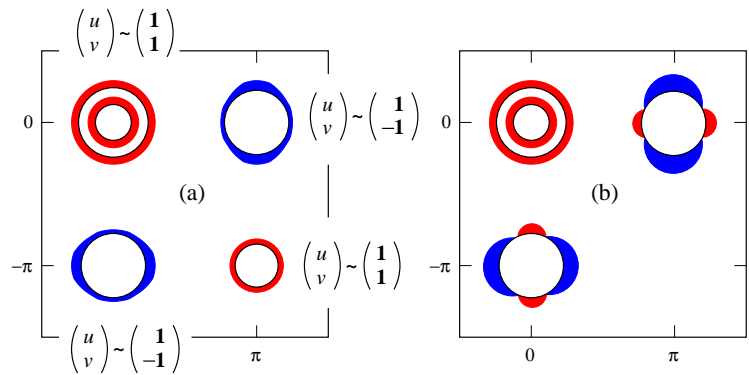}
\caption{(Color online)
(a) Schematic representation of the {\spm} pairing form factor, 
and the associated quasiparticle Nambu spinors. 
Solid black circles represent the normal state Fermi surfaces.
Thickness of colored region around Fermi surfaces indicate magnitude of 
the gap $|\Delta(\v k)|$. Blue means positive and red means negative. 
The two component column vectors are the quasiparticle spinor 
wave functions associated with different Fermi surfaces. 
(b) A schematic representation of the pairing form factor for LaFePO 
obtained by FRG. 
There are four nodes of gap around each electron Fermi surface. 
}
\label{spinor}
\end{center}
\end{figure*}

In a recent paper\cite{mech} the present authors proposed 
the second type of experiment - STM ``quasiparticle-inteference'' experiment. 
This experiment can also probe the relative (spinor) phase between 
the quasiparticles on the electron and hole pockets. Here we repeat 
the argument provided in Ref.~\cite{mech}. 
If the electron and hole Fermi surfaces have out-of-phase order parameters, 
the associated quasiparticle Nambu spinor will be orthogonal 
(left panel, \Fig{spinor}).

For example, under the gauge where the order parameter is real, one of them 
will be $\sim (1,1)$, and the other $\sim (1, -1)$. The matrix elements of 
impurity-induced quasiparticle scattering from the hole to 
the electron Fermi surfaces (or vice versa) is non-zero/zero due to 
a scalar/magnetic impurity 
(which acts as the Pauli matrix $\sigma_z/I_{2\times 2}$ in the Nambu space). 
The reverse is true for the scattering between two hole or two electron Fermi surfaces.  
As a result, if the main scattering source are scalar impurities, 
the quasiparticle interference peaks associated with $(\pm\pi,\pm\pi)$ 
(which arise from the scattering between two electron or two hole pockets, 
note that throughout this paper the ``unfolded'' zone notation of momentum space will be used) 
will be suppressed for bias close to the superconducting gap. 
In contrast, for magnetic impurity scattering, 
the quasiparticle interference peak around $(\pm\pi,0)$ and $(0,\pm\pi)$ 
(which correspond to the scattering between an electron and a hole pocket) 
will be suppressed. 
The contrast between these two sets of quasiparticle interference peaks should 
diminish as the bias is increased/decreased from the gap because 
the spinor phase of the quasiparticle wave functions are 
no longer solely determined by the gap parameter. 
The quasiparticle interference idea \cite{WangLee} has been developed into
a fruitful spectroscopy of studying the cuprates \cite{Hoffman}. 
It has also been used as a method to infer the pairing symmetry by 
Hanaguri \etal \cite{HanaguriCuprate}.
In a recent beautiful STM experiment\cite{Hanaguri} done on 
the Fe$_{1+x}$(Se,Te) compounds, 
the authors control the relative degree between scalar and 
magnetic impurity scattering by changing the density of 
superconducting vortices (which act as magnetic scattering centers). 
Interestingly the above behavior is observed.

Combining the neutron and the STM experiment it is fair to say that 
the evidences for the {\spm} pairing is strong.  
The purpose of the present paper is to tie up the loose ends on 
the theoretical side. 
We study the pairing form factors for LaFePO and Fe(Se,Te) systems. 
These systems exhibit superconductivity without charge doping, 
and are presumably less prone to disorder effects. 
LaFePO, among all iron-based superconductors, shows the strongest evidence 
for the existence of gap nodes\cite{Fletcher,Hicks}. 
For FeSe, there have been contradicting reports of 
nodeless\cite{DongJK} and nodal\cite{Michioka} 
superconductivities, as in many of its pnictides relatives. 
Bulk FeTe is not superconducting but becomes superconductor under 
tensile stress as thin films\cite{HanY}.

We apply the functional renormalization group (FRG) method to study 
the pairing form factor. In particular we incorporate 
the realistic band structures of these materials in the form of tight binding models. 
We then transform the real space on-site interaction into 
two-particle scattering vertex functions defined in the band eigenfunction basis.  
It turns out that much of the differences between 
the pairing form factors of LaFePO, Fe(Se,Te) and the LaFeAsO systems originate from 
the differences in the Fermi surface topology as well as 
the orbital contents of the band wave functions. 
Thus the realistic band structure is indispensable for our purposes.

As pointed out in Ref.~\cite{Scalapino2,Zhai}, 
the scattering vertex function acquires 
important angle dependence around the Fermi pockets due to the change in 
the orbital content of the band wave functions around each Fermi surfaces, 
and this can induce strong variation in 
the {\spm} gap function around the electron pockets. 
Recently it has also been shown that a particular type of angular dependent 
scattering amplitude can even lead to nodal {\spm} form factor\cite{Kuroki2, Chubukov2, Thomale}.

The Hamiltonian we start with is given by the sum of the band structure part and 
the interaction part $H=H_0+H_{int}$.
Because the relevant bands are mostly iron in character, 
we follow the literature in using a tight-binding model
consists of only the iron orbitals. 
Moreover we focus on the two dimensional X-Fe-X (X=As, P, Se) trilayer 
in which the Fe form a square lattice. 
For both materials five-orbital tight-binding models (\ref{equ:tightbinding}) 
fitted to the band structure are used,
\begin{equation}
H_0=\sum_{\vecr,s}\sum_{ij}\sum_{a,b} t_{ij}^{ab}c^\dagger_{ias}c_{jbs}
\label{equ:tightbinding}
\end{equation}
where $a,b=1,5$ label the five Wannier d-orbitals ($3z^2-r^2,xz,yz,x^2-y^2,xy$) of Fe. 
Here $s=\uparrow,\downarrow$ labels spin, $i,j$ labels the iron sites. 
In the above $x$ and $y$ refer the the two orthogonal Fe-Fe directions. 
After a staggered definition of the phase of the $xz$ and $yz$ orbitals, 
the tight-binding Hamiltonian can be made to have the translation symmetry 
consistent with one Fe per unit cell. 
In the rest of the paper we shall use this unit cell and 
the associated Brillouin zone
 (which is referred to as the ``unfolded zone'' in the literature).

For $H_{\rm int}$ we only consider the following on-site interactions
\begin{equation}
\begin{split}
H_{\rm int}= &\ \frac{1}{2}\sum_{i}
\sum_{s,s'}
 [ \sum_{a,b} U_{ab}\, c_{i a s}^\dagger c_{i b s'}^\dagger
 c_{i b s'}^\nd c_{i a s}^\nd \\
&
 + \sum_{a\neq b} J_{ab}
:( c_{i a s}^\dagger c_{i b s}^\nd +{\rm h.c})
 ( c_{i b s'}^\dagger c_{i a s'}^\nd +{\rm h.c.}):
 ].
\end{split}
\label{equ:interactions}
\end{equation}
where $::$ means normal ordering, ${\rm h.c.}$ means hermitian conjugate of
the previous term. When the parameters are available, 
we have studied the full-orbital dependent interaction\cite{Arita}. 
However in the rest of the paper we shall focus on the simpler case where
 $U_{aa}=U$, $U_{ab}=U'$ for $a\neq b$, and $J_{ab}=J$ for $a\neq b$. 
For cases where the orbital-dependent interaction parameters are available 
we have checked that the above simplification does not change 
the qualitative nature of the results. 
The functional renormalization group procedure and result analysis exactly 
follow that used in previous papers by the same authors\cite{FW,mech,Zhai}, 
except that we have achieved slightly better momentum resolution 
in the present work.
In the following we present the results for LaFePO, FeSe and FeTe separately.

\begin{figure*}[tbp]
\begin{center}
\includegraphics[angle=0,scale=0.4]{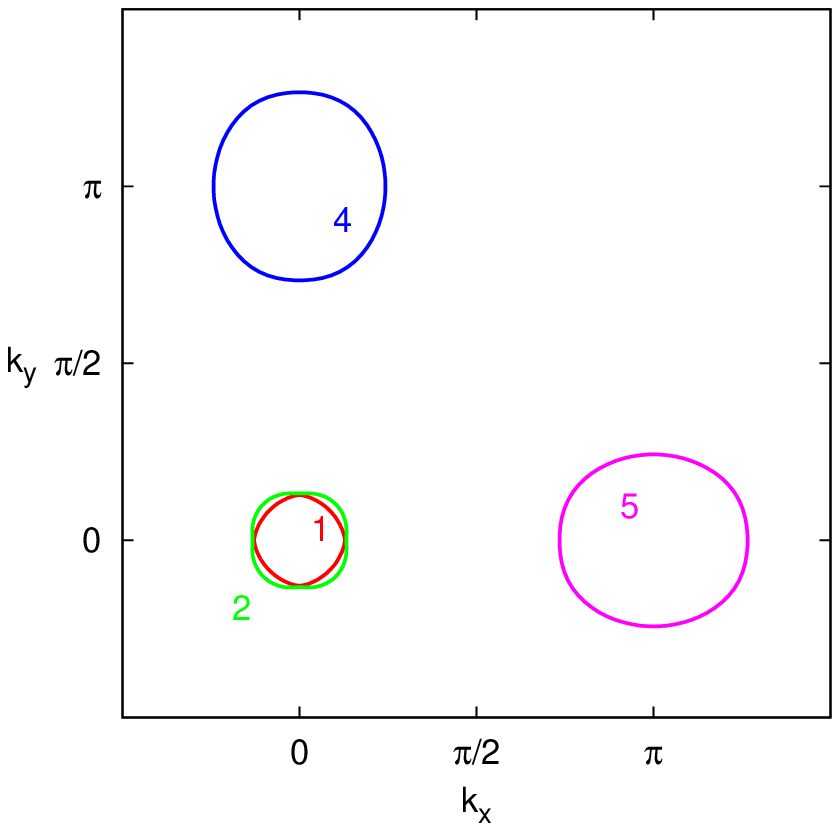}
\includegraphics[angle=0,scale=0.65]{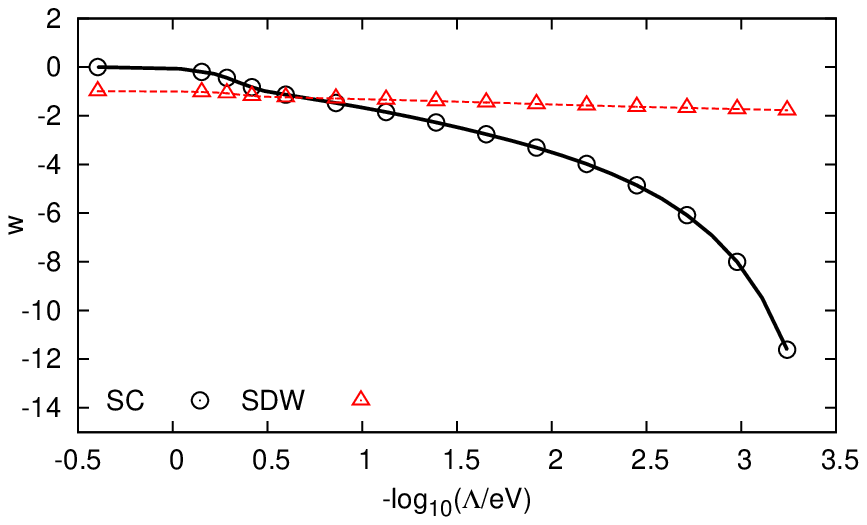}
\includegraphics[angle=0,scale=0.7]{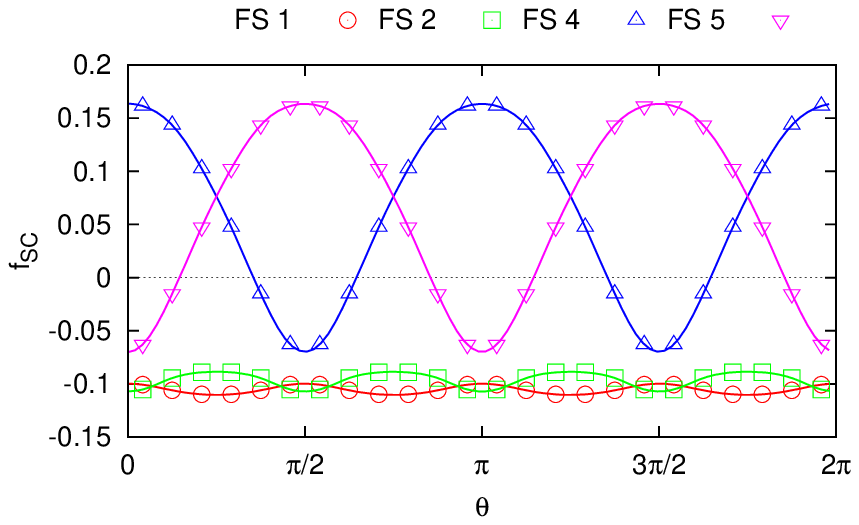}
\caption{(Color online)
Left panel: the $k_z=0$ Fermi surfaces of LaFePO. 
Middle panel: The FRG flow of the interaction strength associated with 
the $(\pi,0)$ SDW and the superconducting pairing. 
$\Lambda$ is the energy cutoff. 
Right panel: the pairing form factor. 
The horizontal axis $\theta$ is the polar angle of Fermi surface points  
with respect to the center of each Fermi surface.  
$\theta=0$ means $+k_x$ direction from the center of the Fermi surface. 
This notation will be used throughout this paper. 
}
\label{LaFePOFS}
\end{center}
\end{figure*}

{\bf LaFePO:} 
For LaFePO there is a discrepancy between the results of 
quantum oscillation experiment\cite{Carrington} and 
the band structure calculations\cite{Kuroki2, LuDH}. 
While the band structure calculations predict the presence of 
a three dimensional $3z^2-r^2$ orbital hole-like Fermi surface centered around
 $\v k=(\pi,\pi,\pi)$, the quantum oscillation experiment did not observe it. 
In our study we follow the quantum oscillation experiment and adapt 
a band structure model where there is only two-dimensional-like Fermi surfaces 
associated with the two hole pockets around $\v k=(0,0)$ and 
two electron pockets around $\v k=(\pi,0)$ and $(0,\pi)$, respectively. 
It is important to note that unlike the hole-doped 122 compounds, 
there is no two-dimensional $xy$ orbital character hole pockets centered at 
 $\v k=(\pi,\pi)$ 
(See \Fig{LaFePOFS}, left panel). 
We shall also comment on why we do not believe the existence of the 
$(\pi,\pi,\pi)$ hole pocket from the theoretical persepctive. 

The FRG flow of the interaction associated with pairing and 
the the $(\pi,0)$ spin density wave (SDW) are shown in the middle panel of 
\Fig{LaFePOFS}. (Strong negative interaction implies ordering instability.)
Despite the fact that at high energies the SDW interaction is more negative, 
the superconducting pairing overwhelms the SDW interaction at low energy cutoffs.
In the right panel of \Fig{LaFePOFS} we present pairing form factors 
associated with the four Fermi surfaces in the left panel. 
We note the following two facts. 
(1) The pairing form factor on each electron Fermi surface has 4 nodes. 
This is schematically represented in the right panel of \Fig{spinor}(b). 
(2) The mean gap function on the electron pockets has opposite sign 
from that on the two $\Gamma$ hole pockets. 
Hence it is justified to call them {\spm} pairing.
The presence of gap nodes on the electron pocket is consistent with 
the observation of the linear-T dependent penetration depth\cite{Fletcher,Hicks}.

\begin{figure*}[tbp]
\begin{center}
\includegraphics[angle=0,scale=0.4]{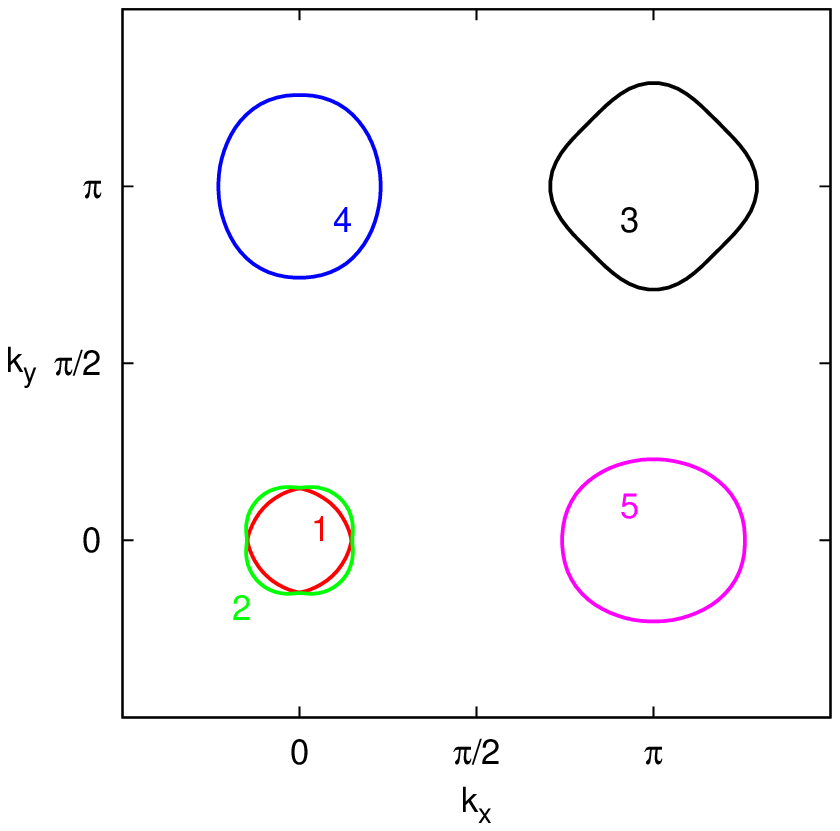}
\includegraphics[angle=0,scale=0.65]{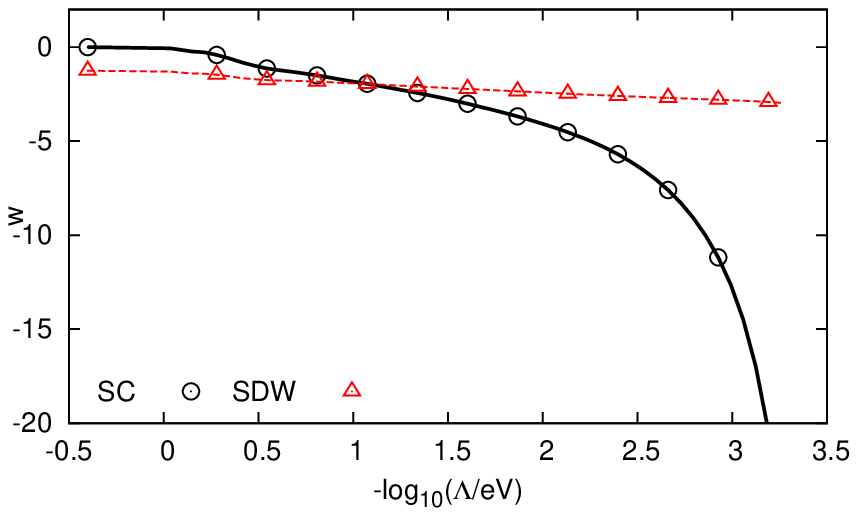}
\includegraphics[angle=0,scale=0.7]{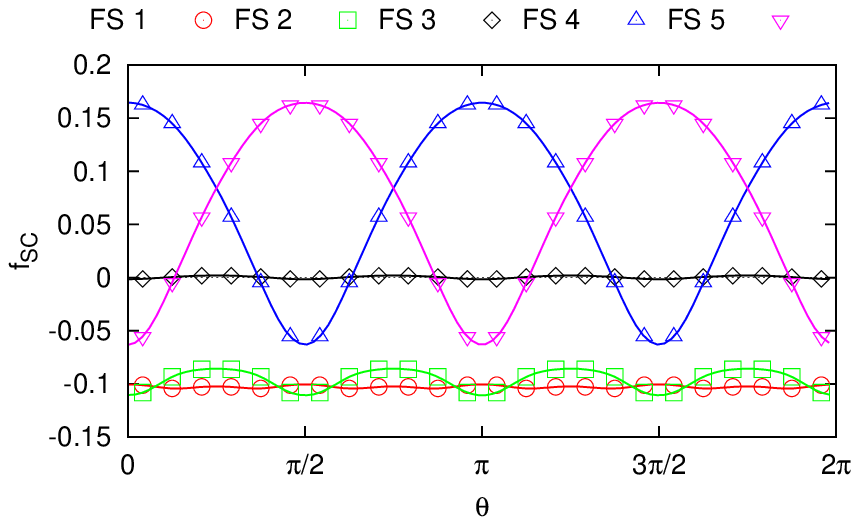}
\caption{(Color online)
Left panel: the Fermi surfaces of LaFePO at $k_z=\pi$. 
Middle panel: The FRG flow of the interaction strength associated with 
the $(\pi,0)$ SDW and the superconducting pairing.
Right panel: the pairing form factor determined by FRG.}
\label{LaFePOFS2}
\end{center}
\end{figure*}

Next we address the issue of the absence/presence of 
the 3D hole pocket around $(\pi,\pi,\pi)$, 
{i.e.}, the conflict between the quantum oscillation experiment and
 the band structure results. 
A cross section of the band structure calculation results at $k_z=\pi$ is 
shown in the left panel of \Fig{LaFePOFS2}. 
The Fermi surface \#3 is the controversial hole pockets. 
It is primarily made up of the $3z^2-r^2$ orbital. 
The rest of Fermi surfaces are connected to those at $k_z=0$ without 
 significant dispersion.
Our FRG result shows that while the bare pair scattering between 
the electron and the $3z^2-r^2$-like hole pocket is appreciable, 
as the energy cutoff is progressively lowered, such pair scattering diminishes. 
This is shown in the right panel of \Fig{LaFePOFS2}, 
where the black diamond symbols mark the form factor on the \#3 Fermi surface - 
it is vanishingly small\cite{private}. 
Clearly if this ungapped Fermi surface exists, it would change 
the temperature dependence of the penetration depth entirely. 
Thus we support the quantum oscillation results and believe that 
the $(\pi,\pi,\pi)$ hole pocket is absent in LaFePO. 
This makes the band structure of LaFePO essentially two dimensional 
like the other 1111 and 122 compounds.

\begin{figure*}[tbp]
\begin{center}
\includegraphics[angle=0,scale=0.4]{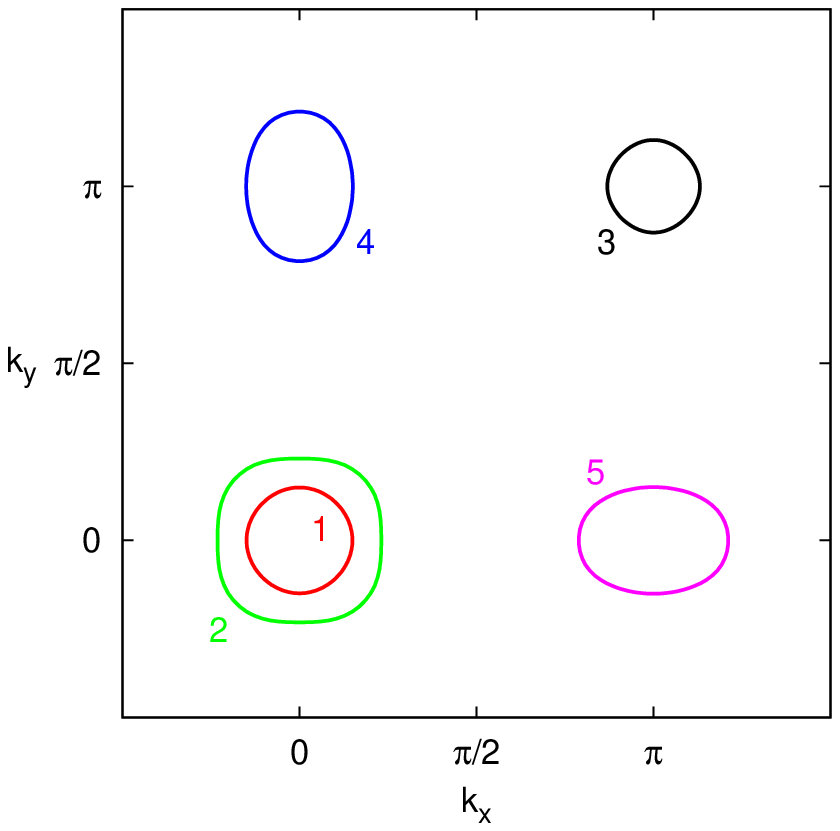}
\includegraphics[angle=0,scale=0.65]{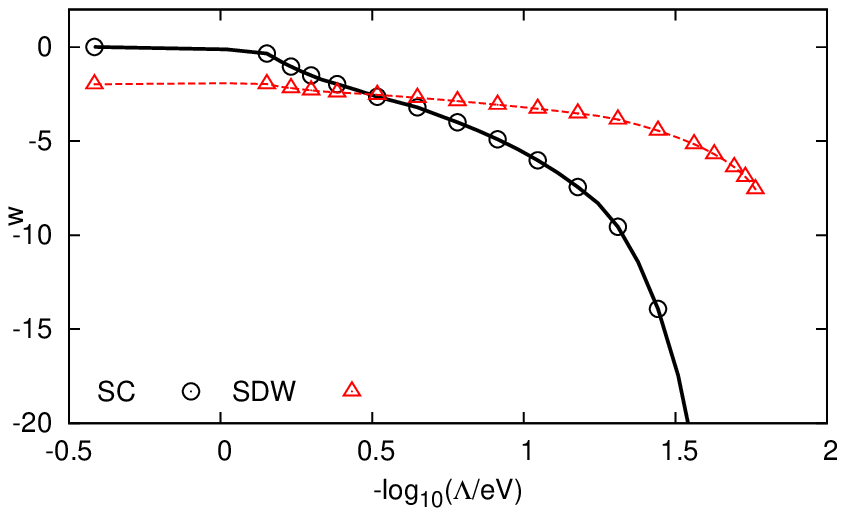}
\includegraphics[angle=0,scale=0.7]{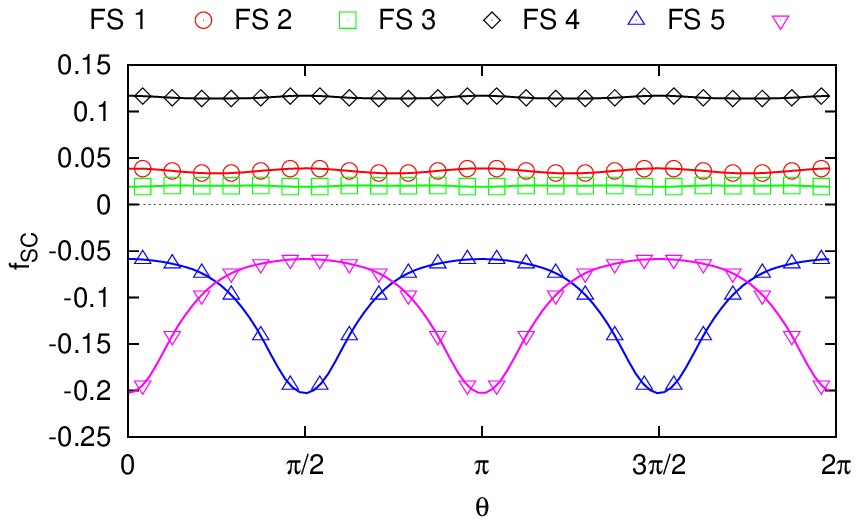}
\caption{(Color online) 
Left panel: the Fermi surfaces of FeSe. 
Middle panel: The FRG flow of the interaction strength associated with 
the $(\pi,0)$ SDW and the superconducting pairing.
Right panel: the pairing form factor.}
\label{FeSeflow}
\end{center}
\end{figure*}

{\bf FeSe:}
Experimental samples of FeSe usually contain small amount of excess Fe. 
But they are experimentally shown as non-essential or even destructive to 
superconductivity\cite{Cava}, and will be ignored in our study. 
The Fermi surface of the FeSe tight-binding model is shown in \Fig{FeSeflow}.
Like the doped 122 systems there are two hole pockets centered around $\v k=(0,0)$, 
two electron pockets around $\v k=(\pi,0),(0,\pi)$, 
and one hole pocket around $\v k=(\pi,\pi)$. 

The FRG flow of the interaction associated with pairing and 
the $(\pi,0)$ spin density wave (SDW) are shown in the middle panel of \Fig{FeSeflow}. 
Again, while the SDW interaction is stronger (more negative) at high energies, 
it is surpassed by the superconducting pairing as the energy cutoff is lowered. 
In the right panel of \Fig{FeSeflow} we present pairing form factors associated with 
the five Fermi surfaces in \Fig{FeSeflow}. We note the following facts. 
(1) The Fermi surface shapes and orbital contents(not shown) are very similar 
to that of LaFeAsO. 
(2) The gap function on the two central hole pockets is quite small compared with 
that on the $(\pi,\pi)$ hole pocket (mainly $xy$ orbital) 
and the electron pockets. 
This suggests the main pairing source is the Cooper scattering between 
the $(\pi,\pi)$ hole pocket and the electron pockets. 
(3) The gap on the electron pockets has substantial variation similar to
our previous results for the 1111 systems\cite{FW}. 
However in this case the large gap is concentrated in regions with dominant 
$xy$ orbital content, i.e. 
the portion of electron Fermi surfaces facing the central hole pockets. 
This is different from our previous result of LaFeAsO where 
the $xy$ orbital dominant part of Fermi surfaces has 
minimal gap.
(4) The gap function takes on opposite sign between the electron and hole pockets. 
Hence the pairing form factor is {\spm}. 
This is consistent with STM quasiparticle interference results of 
 Hanaguri \etal\cite{Hanaguri}.


{\bf FeTe:}
Bulk FeTe is not superconducting and shows 
a different antiferromagnetic(AFM) order than 
the parent iron pnictides\cite{BaoW}. 
Its optical conductance does not show a clear Drude peak and 
does not exhibit a significant change across magnetic transition\cite{WangNLFeTe}. 
The magnetic moment in AFM state is large, about two Bohr magneton\cite{BaoW}. 
All these facts indicate that FeTe may be very different from FeSe 
and iron pnictides, and probably more strongly correlated. 
However
 upon Se substitution of Te superconductivity appears. 
Interestingly, in those superconducting samples neutron scattering showns 
broadened $(\pi,0)$ and $(0,\pi)$ peaks similar to those of the 122 family. 
This provides circumstantial evidences that the $(\pi,0)$ and $(0,\pi)$ 
magnetic scattering is tied to the superconducting pairing. 
Recently superconductivity in FeTe thin films 
under tensile stress was also reported\cite{HanY} 
without Se substitution. 
Knowing the magnetic properties of the superconducting Fe(Se,Te) systems 
it is reasonable to expect that these film to exhibit $(\pi,0)$, $(0,\pi)$ 
antiferromagnetic correlation rather than the $(\pi/2,\pi/2)$ 
antiferromagnetic correlation in bulk FeTe samples. 
Since the $(\pi,0)$, $(0,\pi)$ antiferromagnetic scattering is natural from 
the band structure point of view, 
it is reasonable to start from the itinerant picture when studying 
the superconductivity in FeTe films. 

\begin{figure*}[tbp]
\begin{center}
\includegraphics[angle=0,scale=0.4]{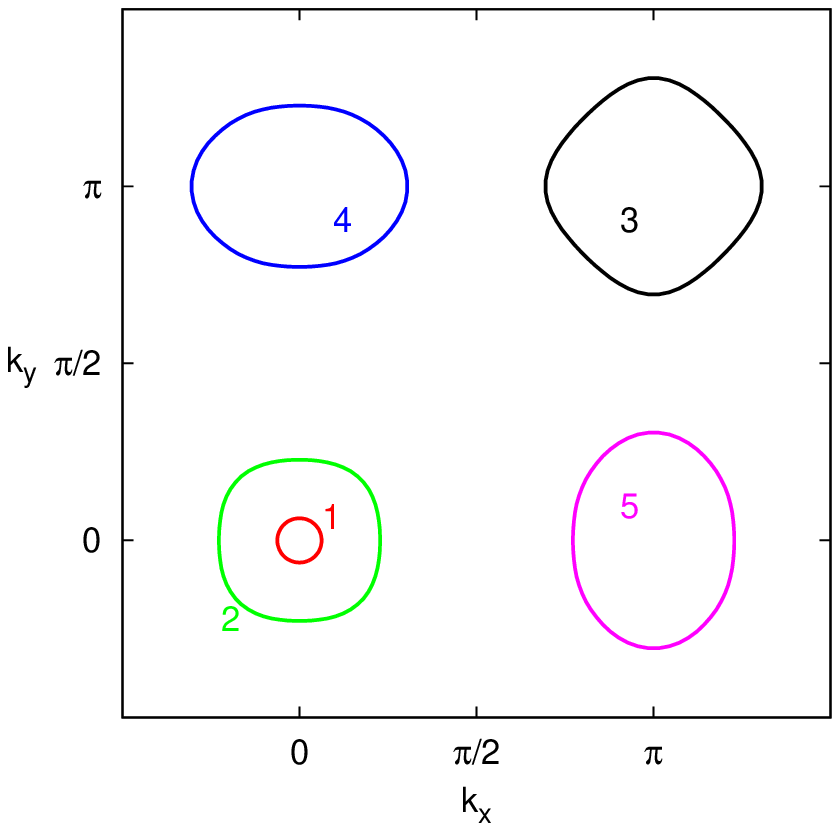}
\includegraphics[angle=0,scale=0.65]{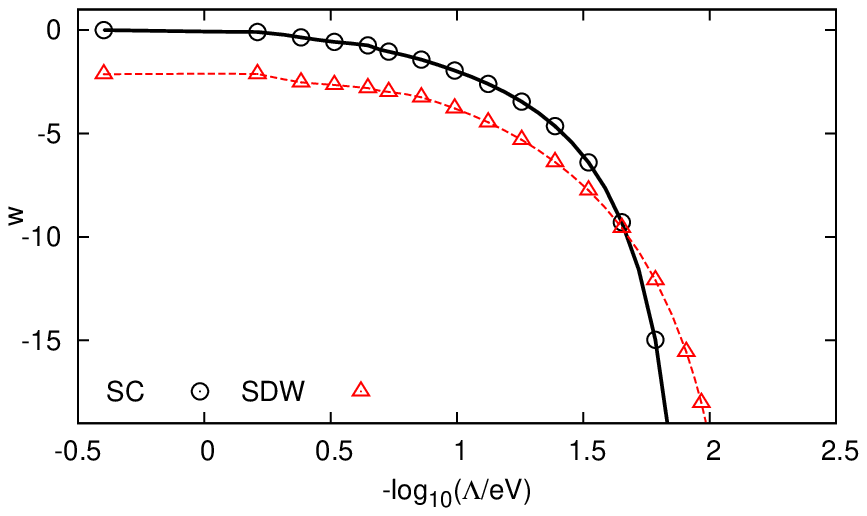}
\includegraphics[angle=0,scale=0.7]{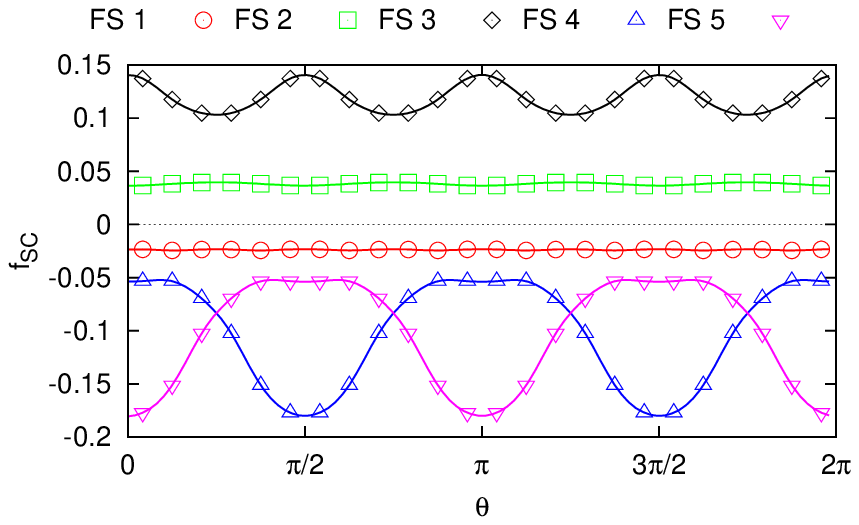}
\caption{(Color online)
Left panel: Fermi surfaces of FeTe. 
Middle panel: The FRG flow of the interaction strength associated with 
the $(\pi,0)$ SDW and the superconducting pairing. 
Right panel: the pairing form factor determined by FRG.}
\label{fig:FeTe}
\end{center}
\end{figure*}

We use the tight-binding model fitted to the 
band structure calculations for FeTe. The band structure has been 
partly confirmed by ARPES measurement\cite{Hsieh}. 
Although it does not capture the unusual magnetism in bulk FeTe, 
we hope it is an appropriate starting point to study the superconductivity 
in the FeTe thin films. 
Excess iron exists in experimental samples of FeTe as in the case of FeSe, 
but will be ignored in our study. 
The Fermi surface, FRG flow and pairing form factor are presented 
in \Fig{fig:FeTe}. 
As expected our results only show the $(\pi,0)$ SDW correlation, 
not the $(\pi/2,\pi/2)$ AFM in bulk FeTe.

We note the following facts:
(1) The Fermi surface shapes and their orbital contents(not shown) are 
very similar to that of LaFeAsO, 
except that the small hole pocket at $\Gamma$ is much smaller, and 
has mainly $xy$ orbital character. 
(2) The gap on electron pockets has substantial variations, 
but does not change sign. Like FeSe the large gap is concentrate in regions with 
dominant $xy$ orbital content. 
(3) The gaps on $(\pi,\pi)$ hole pocket is large and has opposite sign from 
that on the electron pockets. 
Like FeSe, this suggests the root of pairing rests on the Cooper scattering 
between the $(\pi,\pi)$ hole pocket and the electron pockets.
(4) Interestingly, the gaps on the two central hole pockets are 
relatively small and have opposite sign. 
This suggests that these hole pockets play a secondary role in 
the superconducting pairing. The fact that the gap function reverse sign on 
the central hole pockets is a new feature that is absent in 
other iron based superconducting compounds.  

{\bf Real Space Representation of the Pairing Form Factors.}
Our FRG method can only calculate the pairing form factors 
around Fermi surfaces, 
and the pairing order parameter is obtained in band eigenbasis. 
To gain more intuitive picture, it is useful to have a real space picture 
of the pairing.

\begin{figure*}[tbp]
\includegraphics[scale=0.8]{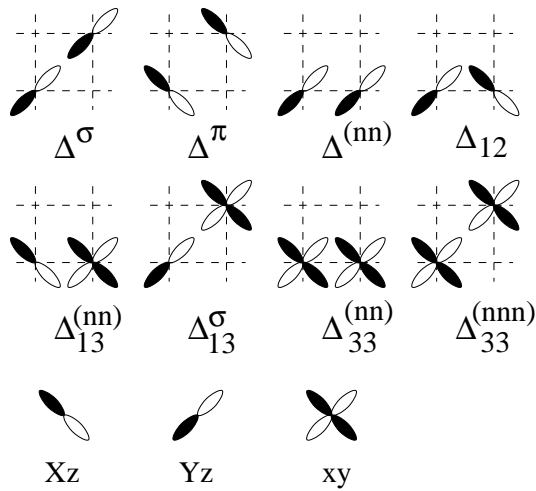}
\caption{
Pictorial representation of the fitting parameters and 
the three orbitals used. 
Empty and filled lobes in the orbital pictures indicate 
$\pm$ sign/symmetry of wave functions. 
Each of the fitting parameters, $\Delta$s, 
represents spin-singlet pairing between the two orbitals shown 
in the corresponding picture.
}
\label{fig:orbitals}
\end{figure*}

Based on our experience and other theoretical works, we will focus on only 
the three $t_2$ orbitals $Xz,Yz,xy$ 
(here capital $X,Y$ refer to the Fe-As directions, namely proper crystal axis) 
in the pairing order parameter. 
Off-site pairing between these three orbitals up to second neighbor on 
the Fe square lattice is included. 
According to the lattice symmetry 
the pairing order parameter has the following form,
\begin{equation}
\Psi_{-\vk,\downarrow}^T
\begin{pmatrix}
\Delta_{11}(k_x,k_y) & \Delta_{12} & \Delta_{13}(k_x,k_y) \\
\Delta_{12} & \Delta_{11}(k_y,-k_x) & \Delta_{13}(k_y,-k_x) \\
-\Delta_{13}(k_x,k_y) & -\Delta_{13}(k_y,-k_x) & \Delta_{33} 
\end{pmatrix}
\Psi_{\vk,\uparrow}
\end{equation}
where $\Psi_{-\vk,s}^T=\begin{pmatrix}
\im c_{Xz,-\vk,s} & \im c_{Yz,-\vk,s} & c_{xy,-\vk,s}
\end{pmatrix}$, 
superscript $^T$ means transpose, 
$s=\uparrow,\downarrow$ labels the spin, 
$\im=\sqrt{-1}$, 
and the entries of the matrix are
\begin{widetext}
\begin{equation*}
\begin{split}
& 
\Delta_{11}(k_x,k_y) = \Delta^{\sigma}\cos(k_x-k_y)
   +\Delta^{\pi}\cos(k_x+k_y)+\Delta^{\rm (nn)}(\cos k_x+\cos k_y), \quad
\Delta_{12} = -\Delta_{12}(\cos k_x -\cos k_y) 
\\ &
\Delta_{13}(k_x,k_y) = \Delta_{13}^{\rm (nn)}(\sin k_x - \sin k_y) 
+\Delta_{13}^{\sigma} \sin(k_x-k_y), \quad
\Delta_{33} =  \Delta_{33}^{\rm (nn)}(\cos k_x + \cos k_y)
+\Delta_{33}^{\rm (nnn)}\cos k_x \cos k_y
\end{split}
\end{equation*}
\end{widetext}
The pictorial representation of the eight fitting parameters 
$\Delta^{\sigma}$, $\Delta^{\pi}$, $\Delta^{\rm (nn)}$, $\Delta_{12}$,
$\Delta_{13}^{\rm (nn)}$, $\Delta_{13}^{\sigma}$, $\Delta_{33}^{\rm (nn)}$,
and $\Delta_{33}^{\rm (nnn)}$ are illustrated in \Fig{fig:orbitals}.

We then project this order parameter onto the Fermi surfaces and compare with 
the FRG results. 
The fitting parameters are obtained by standard least square fit 
and listed in \tbl{tab:fit}. 
A similar real space representation was recently obtained by Kariyado and Ogata 
from RPA results\cite{Kariyado}.
Note that although we have $24$ points around one Fermi surface, 
due to C$_{4v}$ symmetry of our model, 
there are actually only $24/8=3$ independent data points to fit 
per Fermi surface. 
At this stage we have not been able to improve this due to limitation of 
computational power. 
The fitting to LaFePO and FeSe are not very good, with large error estimates. 
This could come from our limited momentum space resolution, or 
that the pairing in these materials actually extend beyond second neighbor. 
So these results should be taken with caution. 

Nonetheless the differences between these materials are clearly visible, 
and are summarized below. 
(1) For LaFeAsO the intra-orbital pairing 
($\Delta^{\sigma}$ and $\Delta^{\rm (nn)}$), between $Xz$-$Xz$ or $Yz$-$Yz$, 
are strongest, therefore the gap is smallest in regions with dominant 
$xy$ character\cite{FW}. 
(2) For LaFePO, the nearest-neighbor pairings, 
inter-orbital $\Delta_{12}$ between $Xz$-$Yz$ and 
intra-orbital pairing $\Delta^{\rm (nn)}_{33}$ 
between $xy$-$xy$, are the strongest, 
which give a nodal $\cos(k_x)+\cos(k_y)$ gap. 
(3) In FeSe the gap is mainly from intra-orbital second neighbor 
$xy$-$xy$ pairing $\Delta^{\rm (nnn)}_{33}$, 
which gives $\cos(k_x)\cos(k_y)$ type nodeless gap. 
The gap variation on electron pockets is mainly due to
the variation of $xy$ orbital content of the wave function. 
Thus it has the opposite anisotropy compared to our LaFeAsO result. 
(4) FeTe is very special, with dominant inter-orbital pairing 
($\Delta_{12}$ and $\Delta^{\rm (nn)}_{13}$) between
$Xz$-$Yz$ and $Xz/Yz$-$xy$. 
The sign structure and variation of gap is a consequence of both this fact and 
the orbital content of the wave functions around Fermi surfaces.

\begin{table*}
\begin{tabular}{|l|llllllll|}
\hline
 &
 $\Delta^{\sigma}$ & $\Delta^{\pi}$ & $\Delta^{\rm (nn)}$ & $\Delta_{12}$ &
 $\Delta_{13}^{\rm (nn)}$ & $\Delta_{13}^{\sigma}$ & $\Delta_{33}^{\rm (nn)}$ &
 $\Delta_{33}^{\rm (nnn)}$ \\
\hline
LaFeAsO $n=6.1$ &
 $0.216(7)$ & $0.085(8)$ & $-0.20(2)$ & $-0.15(1)$ &
 $-0.05(2)$ & $0.021(6)$ & $-0.013(6)$ & $0.03(1)$ \\
LaFePO & 
 $-0.10(7)$ & $0.08(3)$ & $0.08(4)$ & $0.4(1)$ &
 $-0.2(2)$ & $0.09(3)$ & $0.4(4)$ & $0.3(2)$ \\
FeSe & 
 $0.0(2)$ & $0.0(2)$ & $-0.0(2)$ & $-0.00(7)$ &
 $0.07(7)$ & $0.1(1)$ & $0.14(6)$ & $0.4(1)$ \\
FeTe &
 $-0.11(4)$ & $0.14(3)$ & $-0.02(6)$ & $0.31(9)$ & 
 $-0.25(3)$ & $0.12(6)$ & $-0.01(2)$ & $-0.06(1)$ \\
\hline
\end{tabular}
\caption{Fitting parameter for different materials(dopings). 
The row of `LaFeAsO $n=6.1$' is for LaFeAsO model with $6.1$ electrons per site. 
Others are `undoped' ($6$ electrons per site). 
Numbers in bracket are error estimate of the least square fit.
Note that the overall scale and sign are unimportant here.}
\label{tab:fit}
\end{table*}

{\bf Conclusion}
We have so far studied the pairing of LaFeAsO, LaFePO, FeSe and FeTe 
(thin film) using the function renormalization group method. 
The most robust picture that emerge from these studies is that 
pairing is due to the strong positive Cooper scattering between 
the hole and electron Fermi surfaces. The positive value of such scattering is 
responsible for the tendency for the gap function to assume opposite sign on 
the electron and hole pockets. However, on each Fermi surface 
the average magnitude of the gap function, the degrees of its variation as 
a function of angle, and whether it has node or not depend on 
the details of the band wave function as well as the values of 
the local interaction in our model. The present result confirms our earlier 
picture\cite{FW, mech, Zhai} of the antiferromagnetism-triggered pair scattering being 
the mechanism of superconductivity in the iron-based compounds. 
We believe that such mechanism should also at work in systems at the 
brim of band-insulator to semi-metal phase transition 
(for example as a function of pressure). 
On the semi-metal side the charge count requires the electron and hole pockets 
to cover the same area (volume). If the centers of these pockets reside at 
time-reversal invariant $k$-points, 
then intra-pocket pairing can take place. 
If the centers are not located at the time-reversal invariant points, 
then center-of-mass momentum zero Cooper pair will requires 
inter-pocket pairing (time reversal ensure such $k$, $-k$ pockets exists). 
It become increasingly clear with time that 
the highest $T_c$ compounds in the iron-based materials are not 
strongly correlating. This makes our method, 
the function renormalization group an unbiassed, trustworthy method to study 
these systems. 
The neutron resonance and the quasiparticle interference experiments raise 
our hope that perhaps 
after all these years we finally understand an unconventional superconductor. 

We thank Ryotaro Arita, Fengjie Ma and Zhong-Yi Lu for sharing 
the tight-binding fits of their band structure calculations. 
FW is supported by the MIT Pappalardo Fellowship in Physics. 
HZ is supported by the Basic Research Young Scholars Program of  
Tsinghua University, NSFC Grant No. 10944002 and 10847002.
DHL is supported by DOE grant number DE-AC02-05CH11231.


\begin{thebibliography}{}
\bibitem{Mazin} I.I. Mazin, D.J. Singh, M.D. Johannes, and M.H. Du, Phys. Rev. Lett. {\bf 101}, 057003 (2008). 
\bibitem{Kuroki} Kazuhiko Kuroki, Kazuhiko Kuroki, Seiichiro Onari, Ryotaro Arita, Hidetomo Usui, Yukio Tanaka, Hiroshi Kontani, and Hideo Aoki, Phys. Rev. Lett. {\bf 101}, 087004 (2008). 
\bibitem{LiJX}Zi-Jian Yao, Jian-Xin Li, and Z. D. Wang, New J. Phys. {\bf 11}, 025009 (2009). 
\bibitem{HuJP} Kangjun Seo, B. A. Bernevig, and Jiangping Hu, Phys. Rev. Lett. {\bf 101}, 206404 (2008). 
\bibitem{FW} Fa Wang, Hui Zhai, Ying Ran, Ashvin Vishwanath, and Dung-Hai Lee, Phys. Rev. Lett. {\bf 102}, 047005 (2009). 
\bibitem{Chubukov} A. V. Chubukov, D. V. Efremov, and I. Eremin, Phys. Rev. B {\bf 78}, 134512 (2008). 
\bibitem{ZhangFC}Wei-Qiang Chen, Kai-Yu Yang, Yi Zhou, and Fu-Chun Zhang, Phys. Rev. Lett. {\bf 102}, 047006 (2009). 

\bibitem{nodes}
 Yonglei Wang, Lei Shan, Lei Fang, Peng Cheng, Cong Ren, and Hai-Hu Wen, Supercond. Sci. Technol. {\bf 22}, 015018 (2009); 
 Yusuke Nakai, Kenji Ishida, Yoichi Kamihara, Masahiro Hirano, and Hideo Hosono, J. Phys. Soc. Jpn. {\bf 77}, 073701 (2008); 
 H.-J. Grafe, D. Paar, G. Lang, N. J. Curro, G. Behr, J. Werner, J. Hamann-Borrero, C. Hess, N. Leps, R. Klingeler, and B. Buechner, Phys. Rev. Lett. {\bf 101}, 047003 (2008); 
 R. T. Gordon, N. Ni, C. Martin, M. A. Tanatar, M. D. Vannette, H. Kim, G. Samolyuk, J. Schmalian, S. Nandi, A. Kreyssig, A. I. Goldman, J. Q. Yan, S. L. Bud'ko, P. C. Canfield, and R. Prozorov, Phys. Rev. Lett. {\bf 102}, 127004 (2009). 
\bibitem{fullgap}
 T. Y. Chen, Z. Tesanovic, R. H. Liu, X. H. Chen, and C. L. Chien, Nature {\bf 453}, 1224 (2008); 
 L. Malone, L. Malone, J.D. Fletcher, A. Serafin, A. Carrington, N.D. Zhigadlo, Z. Bukowski, S. Katrych, and J. Karpinski, 	Phys. Rev. B {\bf 79}, 140501(R) (2009); 
 H. Ding, P. Richard, K. Nakayama, T. Sugawara, T. Arakane, Y. Sekiba, A. Takayama, S. Souma, T. Sato, T. Takahashi, Z. Wang, X. Dai, Z. Fang, G. F. Chen, J. L. Luo, and N. L. Wang, Euro. Phys. Lett., {\bf 83}, 47001 (2008); 
 Takeshi Kondo, A. F. Santander-Syro, O. Copie, Chang Liu, M. E. Tillman, E. D. Mun, J. Schmalian, S. L. Bud'ko, M. A. Tanatar, P. C. Canfield, and A. Kaminski, Phys. Rev. Lett. {\bf 101}, 147003 (2008); 
 K. Hashimoto, T. Shibauchi, S. Kasahara, K. Ikada, S. Tonegawa, T. Kato, R. Okazaki, C. J. van der Beek, M. Konczykowski, H. Takeya, K. Hirata, T. Terashima, and Y. Matsuda, Phys. Rev. Lett. {\bf 102}, 207001 (2009). 

\bibitem{Tsuei}C. C. Tsuei, J. R. Kirtley, C. C. Chi, Lock See Yu-Jahnes, A. Gupta, T. Shaw, J. Z. Sun, and M. B. Ketchen, Phys. Rev. Lett. {\bf 73}, 593 (1994).
\bibitem{Harlingen}D. A. Wollman, D. J. Van Harlingen, W. C. Lee, D. M. Ginsberg, A. J. Leggett, Phys. Rev. Lett. {\bf 71}, 2134 (1993).
\bibitem{Moler}C.W. Hicks, T.M. Lippman, M.E. Huber, Z.A. Ren, Z.X. Zhao, and K.A. Moler, J. Phys. Soc. Jpn. {\bf 78}, 013708 (2009). 

\bibitem{Maier}T.A. Maier, and D.J. Scalapino, Phys. Rev. B {\bf 78}, 020514(R) (2008). 
\bibitem{Korshunov} M.M. Korshunov, and I. Eremin, Phys. Rev. B {\bf 78}, 140509(R) (2008). 

\bibitem{Lumsden} M. D. Lumsden, A. D. Christianson, D. Parshall, M. B. Stone, S. E. Nagler, G.J. MacDougall, H. A. Mook, K. Lokshin, T. Egami, D. L. Abernathy, E. A. Goremychkin, R. Osborn, M. A. McGuire, A. S. Sefat, R. Jin, B. C. Sales, and D. Mandrus, Phys. Rev. Lett. {\bf 102}, 107005 (2009). 
\bibitem{ChiS} Songxue Chi, Astrid Schneidewind, Jun Zhao, Leland W. Harriger, Linjun Li, Yongkang Luo, Guanghan Cao, Zhu'an Xu, Micheal Loewenhaupt, Jiangping Hu, and Pengcheng Dai, Phys. Rev. Lett {\bf 102}, 107006 (2009). 
\bibitem{LiS} Shiliang Li, Ying Chen, Sung Chang, Jeffrey W. Lynn, Linjun Li, Yongkang Luo, Guanghan Cao, Zhu'an Xu, and Pengcheng Dai, Phys. Rev. B {\bf 79}, 174527 (2009). 
\bibitem{Inosov} D. S. Inosov, J. T. Park, P. Bourges, D. L. Sun, Y. Sidis, A. Schneidewind, K. Hradil, D. Haug, C. T. Lin, B. Keimer, and V. Hinkov, arXiv:0907.3632 (2009).
\bibitem{Zhao} Jun Zhao, Louis-Pierre Regnault, Chenglin Zhang, Miaoying Wang, Zhengcai Li, Fang Zhou, Zhongxian Zhao, and Pengcheng Dai, arXiv:0908.0954 (2009).

\bibitem{mech} Fa Wang, Hui Zhai, and Dung-Hai Lee, Europhys. Lett. {\bf 85}, 37005 (2009). 

\bibitem{WangLee} Q.-H. Wang, and D.-H. Lee, Phys. Rev. B {\bf 67}, 020511(R) (2003). 

\bibitem{Hoffman}
J. E. Hoffman, K. McElroy, D.-H. Lee, K. M Lang, H. Eisaki, S. Uchida, and J.C. Davis, Science {\bf 297}, 1148 (2002); 
K. McElroy, D.-H. Lee, J. E. Hoffman, K. M Lang, J. Lee, E. W. Hudson, H. Eisaki, S. Uchida, and J.C. Davis, Nature (London) {\bf 422}, 592 (2003); 
J. C. Davis, Acta Physica Polonica A {\bf 104}, 193 (2003); 
J. A. Slezak, J. Lee, J. C. Davis, MRS Bulletin, {\bf 30}, 437 (2005);
J. Lee, J. A. Slezak, J. C. Davis, J. Phys. Chem. Solids, {\bf 66}, 1370 (2005);
T. Hanaguri, Y. Kohsaka, J. C. Davis, C. Lupien, I. Yamada, M. Azuma, M. Takano, K. Ohishi, M. Ono, and H. Takagi, Nature Phys. {\bf 3}, 865 (2007); 
Jhinhwan Lee, K. Fujita, A.R. Schmidt, Chung Koo Kim, H. Eisaki, S. Uchida, and J.C. Davis, Science {\bf 325}, 1099 (2009). 
\bibitem{HanaguriCuprate}T. Hanaguri, Y. Kohsaka, M. Ono, M. Maltseva, P. Coleman, I. Yamada, M. Azuma, M. Takano, K. Ohishi, and H. Takagi, Science {\bf 323}, 923 (2009). 
\bibitem{Hanaguri} T. Hanaguri, S. Niitaka, K. Kuroki, and H. Takagi, Science {\bf 328}, 474 (2010).

\bibitem{Fletcher} J.D. Fletcher, A. Serafin, L. Malone, J. Analytis, J-H Chu, A.S. Erickson, I.R. Fisher, and A. Carrington, Phys. Rev. Lett. {\bf 102}, 147001 (2009). 
\bibitem{Hicks}Clifford W. Hicks, Thomas M. Lippman, Martin E. Huber, James G. Analytis, Jiun-Haw Chu, Ann S. Erickson, Ian R. Fisher, and Kathryn A. Moler, Phys. Rev. Lett. {\bf 103}, 127003 (2009). 
\bibitem{DongJK}J. K. Dong, T. Y. Guan, S. Y. Zhou, X. Qiu, L. Ding, C. Zhang, U. Patel, Z. L. Xiao, and S. Y. Li, Phys. Rev. B {\bf 80}, 024518 (2009). 
\bibitem{Michioka}Chishiro Michioka, Hiroto Ohta, Mami Matsui, Jinhu Yang, Kazuyoshi Yoshimura, and Minghu Fang, arXiv:0911.3729 (2009). 
\bibitem{HanY} Y. Han, W. Y. Li, L. X. Cao, X. Y. Wang, B. Xu, B. R. Zhao, Y. Q. Guo, and J. L. Yang, Phys. Rev. Lett. {\bf 104}, 017003 (2010). 

\bibitem{Scalapino2}T.A. Maier, S. Graser, D.J. Scalapino, and P.J. Hirschfeld, Phys. Rev. B {\bf 79}, 224510 (2009).
\bibitem{Zhai} Hui Zhai, Fa Wang, and Dung-Hai Lee, Phys. Rev. B {\bf 80}, 064517 (2009). 
\bibitem{Kuroki2}Kazuhiko Kuroki,Hidetomo Usui, Seiichiro Onari, Ryotaro Arita, and Hideo Aoki, Phys. Rev. B {\bf 79}, 224511 (2009). 
\bibitem{Chubukov2} A. V. Chubukov, M. G. Vavilov, and A. B. Vorontsov, Phys. Rev. B {\bf 80}, 140515(R) (2009). 
\bibitem{Thomale} R. Thomale, C. Platt, J. Hu, C. Honerkamp, and B. A. Bernevig, Phys. Rev. B {\bf 80}, 180505(R) (2009). 
\bibitem{Arita}Kazuma Nakamura, Ryotaro Arita, and Masatoshi Imada, {J. Phys. Soc. Jpn.} {\bf 77}, {093711} {(2008)}
\bibitem{Carrington} A. Carrington, A. I. Coldea, J. D. Fletcher, N. E. Hussey, C. M. J. Andrew, A. F. Bangura, J. G. Analytis, J.-H. Chu, A. S. Erickson, I. R. Fisher, and R. D. McDonald, Physica C {\bf 469}, 459 (2009). 
\bibitem{LuDH} D. H. Lu, M. Yi, S.-K. Mo, A. S. Erickson, J. Analytis, J.-H. Chu, D. J. Singh, Z. Hussain, T. H. Geballe, I. R. Fisher, and Z.-X. Shen, Nature {\bf 455}, 81 (2008). 
\bibitem{private} The same conclusion is obtained by the authors of Ref.~\cite{Kuroki2} (private communication). 
\bibitem{Cava}A. J. Williams, T. M. McQueen, and R. J. Cava, Sol. St. Comm. {\bf 149}, 1507 (2009). 
\bibitem{BaoW}Wei Bao, Y. Qiu, Q. Huang, M.A. Green, P. Zajdel, M.R. Fitzsimmons, M. Zhernenkov, M. Fang, B. Qian, E.K. Vehstedt, J. Yang, H.M. Pham, L. Spinu, and Z.Q. Mao, Phys. Rev. Lett. {\bf 102}, 247001 (2009). 
\bibitem{WangNLFeTe}G. F. Chen, Z. G. Chen, J. Dong, W. Z. Hu, G. Li, X. D. Zhang, P. Zheng, J. L. Luo, and N. L. Wang, Phys. Rev. B {\bf 79}, 140509(R) (2009). 
\bibitem{Hsieh}Y. Xia, D. Qian, L. Wray, D. Hsieh, G. F. Chen, J. L. Luo, N. L. Wang,
and M. Z. Hasan, Phys. Rev. Lett. 103, 037002 (2009). 
\bibitem{Kariyado}Toshikaze Kariyado, Masao Ogata, 	J. Phys. Soc. Jpn. {\bf 79}, 033703 (2010). 

\end{thebibliography}
\end{document}